\documentstyle[12pt,epsfig,multirow]{article}

\newcommand{\be}{\begin{eqnarray}}
\newcommand{\ee}{\end{eqnarray}}

\def\nue{{\nu_e}}
\def\anue{{\bar\nu_e}}
\def\numu{{\nu_{\mu}}}

\newcommand{\ms}{\Delta m^2_{21}}
\newcommand{\ma}{\Delta m^2_{31}}

\newcommand{\sss}{\sin^2 \theta_{12}}

\newcommand{\stch}{\sin^2 2\theta_{13}}

\newcommand{\sta}{\sin^22 \theta_{23}}

\newcommand{\stcht}{\sin^2 2\theta_{13}{\mbox {~(true)}}}

\newcommand{\dcpt}{\delta_{\mathrm{CP}}{\mbox {~(true)}}}
\newcommand{\dcp}{\delta_{\mathrm{CP}}}

\newcommand{\sig}{$3\sigma$}

\newcommand{\br}{$^8$B~}
\newcommand{\li}{$^8$Li~}
\newcommand{\he}{$^6$He~}
\newcommand{\neon}{$^{18}$Ne~}

\def\gtap{\ \raisebox{-.4ex}{\rlap{$\sim$}} \raisebox{.4ex}{$>$}\ }

\newcommand{\stheta}{\sin^22\theta_{13}}

\begin{document}

\thispagestyle{empty}
\vskip -10pt
\begin{flushright}
\texttt{HRI-P-08-04-001}\\
\texttt{CU-PHYSICS/08-2008}\\
\texttt{HRI-RECAPP-08-04}\\
\end{flushright}
\vskip 20pt

\begin{center}
{\Large \bf Exceptional Sensitivity to Neutrino Parameters with a Two-Baseline
Beta-Beam Set-up} 

\vspace{.4in}

{\bf {Sanjib Kumar Agarwalla$^{\star,\dagger,a}$, 
Sandhya Choubey$^{\star,b}$, Amitava Raychaudhuri$^{\star,\dagger,c}$}} 

\vskip 0.4cm

$^\star${\normalsize \it Harish-Chandra Research Institute,} \\
{\normalsize \it Chhatnag Road, Jhunsi, Allahabad  211019, India}

\vskip 0.3cm

$^\dagger${\normalsize \it Department of Physics, University of Calcutta,} \\ 
{\normalsize \it 92 Acharya Prafulla Chandra Road, Kolkata  700009, India}
\vskip 1.8cm

{\bf ABSTRACT}

\end{center}

We examine the reach of a Beta-beam experiment
with two detectors at carefully chosen baselines for
exploring neutrino mass parameters.  Locating the source at CERN,
the two detectors and baselines are: (a) a 50 kton iron
calorimeter (ICAL) at a baseline of around 7150 km which is roughly the
magic baseline, e.g., ICAL@INO, and
(b) a 50 kton Totally Active Scintillator Detector 
at a distance of
730 km, e.g., at Gran Sasso. We choose \br and \li source ions
with a boost factor $\gamma$ of 650 for the magic baseline while
for the closer detector we consider $^{18}$Ne and $^6$He ions with a
range of Lorentz boosts. We find that the locations of the two
detectors complement each other leading to an exceptional high
sensitivity. 
With $\gamma=650$ for $^8$B/$^8$Li and  $\gamma=575$ for 
$^{18}$Ne/$^6$He and   
total luminosity corresponding to $5\times (1.1
\times 10^{18})$ and $5\times (2.9\times 10^{18})$ useful ion
decays in neutrino and antineutrino modes respectively, we  find
that the two-detector set-up can probe 
maximal CP violation and establish 
the neutrino mass ordering if $\sin^22\theta_{13}$ is $1.4
\times 10^{-4}$ and $2.7 \times 10^{-4}$, respectively, or more.
The sensitivity reach for  $\sin^22\theta_{13}$ itself is $5.5
\times 10^{-4}$.
With a factor of 10 higher luminosity, the corresponding 
$\sin^22\theta_{13}$ reach of this set-up 
would be $1.8 \times 10^{-5}$, 
$4.6 \times 10^{-5}$ and $5.3 \times 10^{-5}$ respectively 
for the above three performance indicators. 
CP violation can be discovered for 64\% of 
the possible $\dcp$ values for 
$\sin^22\theta_{13} \geq 10^{-3}$ ($\geq 8\times 10^{-5}$), for 
the standard luminosity (10 times enhanced luminosity). 
Comparable physics performance can be achieved in a 
set-up where data from CERN to INO@ICAL is combined with 
that from CERN to the Boulby mine in United Kingdom,
a baseline of 1050 km.

\vskip 1.5cm

\noindent $^a$ email: sanjib@hri.res.in

\noindent $^b$ email: sandhya@hri.res.in 

\noindent $^c$ email: raychaud@hri.res.in

\newpage

\section{Introduction}

Long baseline experiments with powerful neutrino beams 
\cite{t2k,nova,iss,issphysics} from 
upgraded accelerator facilities are the next frontier 
in neutrino oscillation physics. Set-ups with intense neutrino beam 
sources and smart detector technologies are being planned to 
fathom the hitherto uncharted regimes of the neutrino 
mass matrix. In particular, the long baseline experiments 
are being designed to measure the third mixing angle $\theta_{13}$, 
CP phase $\dcp$ and $sgn(\ma)$ {\it aka}, 
the neutrino mass hierarchy\footnote{In this paper we 
define $\Delta m^2_{ij} = m_i^2 - m_j^2$ and refer to 
$sgn(\ma)$ as the neutrino mass hierarchy -- 
$sgn(\ma) >0$ is called ``normal hierarchy''(NH) 
while $sgn(\ma) <0$ is called ``inverted hierarchy''(IH). 
We stress that the arguments are valid for both hierarchical 
as well as quasi-degenerate neutrino mass spectra.}.
The most 
promising avenue for this purpose is the $\nue \rightarrow \numu$
oscillation channel $P_{e\mu}$  
(or its T-conjugate $P_{\mu e}$),  
often referred to in the literature as the 
``golden channel'' \cite{golden}. It was realized that 
while  this channel can be used in the most 
cost-effective way to measure all three neutrino 
oscillation parameters listed above, it is also rife 
with the so-called problem of ``parameter degeneracies''. 
These are identified as the 
($\theta_{13},\dcp$) intrinsic degeneracy~\cite{intrinsic},
the ($sgn(\ma),\dcp$) degeneracy~\cite{minadeg}, and 
the ($\theta_{23},\pi/2-\theta_{23}$) degeneracy~\cite{th23octant}.
Together they could result in as many as eight-fold degenerate 
solutions ~\cite{eight}, of which, obviously, only one is true. 
This is evidently a very undesirable situation and a large 
body of existing literature is devoted to finding ways of 
combating this menace 
\cite{intrinsic,diffLnE,t2ksimulation,silver,dissappear,pee,addatm,
cernmemphys,addreact}. A particularly attractive way of 
killing the clone solutions from the $\dcp$ dependent 
degeneracies is to perform the experiment at the 
``magic baseline'' \cite{magic,magic2,petcov}. At this 
baseline all $\dcp$ dependent terms drop out, 
providing an ideal bedrock for measuring $\theta_{13}$ and 
$sgn(\ma)$. 

In \cite{paper1,betaino1,betaino2} we expounded 
the sensitivity reach of a 
magic baseline experimental set-up where the neutrino 
source is a Beta-beam \cite{zucc} at CERN and the detector is a 
large magnetized iron calorimeter (ICAL) at the India-based
Neutrino Observatory\footnote{ The CERN to INO distance is 7152
km, which is tantalizingly close to the magic baseline.} (INO)
\cite{ino}.  Using a 50 kton fiducial mass and 80\% detector
efficiency\footnote{We give details about the detector
configuration in section 3.} for ICAL@INO, and considering a
Beta-beam using \br and \li as ion source \cite{rubbia,mori},
with a Lorentz boost $\gamma=650$ and assuming $5\times (1.1
\times 10^{18})$ and $5\times (2.9\times 10^{18})$ useful ion
decays in neutrino and antineutrino modes respectively, we showed
that this set-up could unambiguously probe $sgn(\ma)$ at
$3\sigma$ if\footnote{These are the values of the parameters
chosen by Nature, to be distinguished from the fitted values.
Throughout this paper we denote the true value of a parameter by
putting ``(true)'' after the symbol for the parameter.}  $\stcht > 5.6
\times 10^{-4}$ and find a signal for $\theta_{13}$ at the same
level of significance if $\stcht
> 5.1 \times 10^{-4}$, {\it independent of the true neutrino mass
hierarchy and} $\dcpt$.  The $\theta_{13}$ and hierarchy
sensitivity reach of this ``magical'' set-up is therefore
superior to that of most other rival proposals involving
Beta-beams
\cite{cernmemphys,oldpapers,
donini130,doninibeta,newdonini,bc1,bc2,fnal,betaoptim,volpe,
doninialter,boulby,nf07sc}, and is almost comparable to that
possible with a Neutrino Factory \cite{nufactoptim} at the magic
baseline.  This tremendous sensitivity of the CERN-INO Beta-beam
project to $\theta_{13}$ and hierarchy comes from a combination
of being close to the magic baseline as well as from being
sensitive to near-maximal matter effects.

The main drawback of the 
CERN-INO Beta-beam set-up is 
its inability to determine $\dcp$, since being 
almost at the magic baseline it is insensitive to the CP
phase. Note that this is also true 
for the Neutrino Factory experiment at the 
magic baseline. 
This is why the optimal Neutrino Factory 
set-up demands a combination of two baselines, one magic and 
the other around $L=4000$ km, where one has the best 
sensitivity to $\dcp$ \cite{nufactoptim}. 
For Beta-beams, one has to optimize not only 
on $\gamma$, the baseline $L$, and the luminosity, but also on the choice of 
the source ions. 
A detailed optimization study 
has revealed that for intermediate values of $\gamma$,
one would preferentially use 
\br and \li at the magic baseline for the mass hierarchy, 
\neon and \he at an intermediate baseline of $L=600-900$ km 
for CP violation, and either \neon and \he at the intermediate  
baseline or \br and \li at the magic baseline for the 
$\sin^22\theta_{13}$ discovery \cite{bboptim}. 

In this paper we 
consider a two-baseline Beta-beam set-up, one with $L=7152$ km,  
the CERN-INO baseline, and another with $L=730$ km
which is the CERN-Gran Sasso (LNGS) distance. For the CERN-INO 
case  \br and \li are the preferred source ions 
and we take $\gamma=650$.  
For the CERN-LNGS set-up, on the other hand, we choose the \neon
and \he  
ions and allow their $\gamma$ to vary between 250-650\footnote{With 
the existing facilities at CERN, one will be able 
to accelarate $^{6}$He to only about $\gamma=150$. This 
corresponds to a boost factor of about 250 for 
$^{18}$Ne and 280 for $^8$B, due to the different charge 
to mass ratios for these ions. 
However, with the ``Super-SPS'', an
upgraded version of the SPS with super-conducting magnets
\cite{bc2,doninibeta} it should be possible to accelarate
$^{6}$He to $\gamma=350$, which corresponds to  $\gamma=575$ for 
$^{18}$Ne and $\gamma=650$ for $^8$B. The Tevatron at 
Fermilab could produce a Beta-beam with similar boost factors. 
Higher acceleration at CERN would require the use of the 
LHC itself, with $\gamma > 1000$ possible \cite{matsmoriond}.}
in order to examine the dependence of the sensitivity on the value of the 
boost factor. Since the \br and \li ions would produce 
multi-GeV neutrino beams for $\gamma=650$, 
we use a 50 kton iron 
calorimeter (ICAL) for the longer baseline at INO 
and call this set-up CERN-ICAL@INO. 
For the intermediate baseline option, since we are 
interested in the lower energy \neon and \he ions, 
we assume a 50 kton 
Totally Active Scintillator Detector (TASD) in order to 
harness the low energy events required for better CP sensitivity. 
We present results for 
$5\times (1.1 \times 10^{18})$  
and $5\times (2.9\times 10^{18})$ 
useful ion decays 
in neutrino and antineutrino modes respectively for both 
baseline set-ups. We also show the projected sensitivity if 
one achieves statistics that are larger by a factor of 10. 
Since the statistics depend on the product of the 
size of the detector, the exposure time, detector 
efficiency and the 
number of useful ion decays, this one order of magnitude increase 
could come from a combination of enhanced performance in 
any of the factors mentioned above. In particular, 
it might be possible to have 10 times larger useful ion decays 
per year in the storage ring \cite{mats_talk_RAL_2008}. 
ICAL@INO could also be upgraded to 100 kton. 

Note that 
while most of our figures and discussion in the text 
would explicitly be for the CERN-LNGS intermediate 
baseline set-up, we also present results for the 
case where a \neon and \he Beta-beam is shot from 
CERN to the Boulby mine in the United Kingdom 
(see \cite{boulby} and references therein). 
The CERN-Boulby distance is 1050 km, and one can check 
from Fig. 5 of \cite{bboptim} that the CP sensitivity
for this baseline is only marginally 
weaker than that for $L=730$ km. Therefore, when combined 
with the CERN-INO data, we expect 
similar performance for this baseline option as well. 

In some sense this work is a part of an ongoing international
exercise in sharpening the full capability of long baseline
Beta-beam experiments to explore neutrino properties. In addition
to the work discussed earlier in the Introduction, some of the
milestones on this route can be identified as the
CERN-MEMPHYS proposal \cite{cernmemphys,oldpapers,donini130},
its variants with higher $\gamma$ \cite{bc1,bc2}, using a cocktail of
sources \cite{doninialter}, detector options, {\em etc}. 

While the best CP violation sensitivity comes with 
\neon and \he as source at our chosen intermediate baseline, 
it might be more convenient and less demanding to use the 
same set of ions at both the baselines. 
The optimal baseline 
for CP studies with $^8$B and $^8$Li
was seen to be around 1000-2000 km in \cite{bboptim}. 
In \cite{newdonini}, the authors proposed a set-up with this 
scenario where they considered a two-baseline combination 
with a longer $L\simeq 7000$ km and a shorter $L\simeq 2000$ km. 
A $\gamma=350$ Beta-beam produced by \br and \li as source ions 
feeds both detectors and they consider up to 
$5\times (10 \times 10^{18})$ total 
useful ion decays in both neutrino and antineutrino modes. 
They consider 50 kton of magnetized iron as the end detector 
for both baselines. 
In this set-up the mass hierarchy can be discovered if 
$\stcht > 3 \times 10^{-4}$ and CP 
violation can be established for 70\% of the possible 
$\dcpt$  
values if $\stcht > 10^{-3}$. Signal for the 
mixing angle itself could be
observed for $\stcht > 1 \times 10^{-4}$. 
For $5\times (2 \times 10^{18})$ total useful decays, 
which is comparable to our standard luminosity, 
the corresponding reaches are 
$\stcht \gtap 10^{-3}$ for the mass hierarchy, and  
$\stcht \gtap 8 \times 10^{-4}$ for discovering $\theta_{13}$. 
CP violation sensitivity gets severely restricted and deteriorates, 
especially for $\dcpt=270^\circ$.  
The poor CP reach for lower luminosity comes 
due to the choice of \br and \li as 
source ions (see \cite{bboptim} for a detailed discussion). 
Since much better CP 
sensitivity for plausible luminosity and $\gamma$ 
can be achieved with the 
\neon and \he combination, 
we will exclusively use them for our intermediate baseline 
option of CERN to LNGS. Since our chosen 
$\gamma$ is larger, 
we also achieve sensitivity to mass hierarchy and 
$\theta_{13}$ which is better by at least an order of magnitude 
for the same number of useful ions decays. 

The paper is organized as follows. In the next section we briefly
discuss the oscillation probability for the golden channel and
highlight the magic baseline behavior. The experimental set-up
which we consider is outlined in the subsequent section. In the
following section we explore the prospects of the TASD@LNGS
detector, in a stand-alone mode, for exploring CP violation. This
is to set the stage for the next section where the results of a
combined two-baseline analysis are presented with emphasis on the
sensitivities to CP violation, $\theta_{13}$, and the neutrino
mass hierarchy. We end by summarizing the results for 
our combined CERN-INO and CERN-LNGS set-ups and make a 
comparative discussion of this set-up {\it vis-a-vis} 
the combined CERN-INO and CERN-Boulby set-up. We also 
compare our set-up against the one studied in \cite{newdonini} and 
discuss the pros and cons of the two proposals. 
Some remarks about the beam related as well as atmospheric 
neutrino backgrounds are collected in the Appendix.

\section{Golden channel oscillations}

For the results presented in this paper we have calculated the
neutrino 
oscillation probability in matter exactly, using the PREM 
profile for the Earth matter density \cite{prem}. However, 
for elucidating the behavior of  
neutrino oscillations as a function of baseline and/or 
neutrino energy, it is useful to exploit the approximate analytic 
formula for the golden channel probability 
$P_{e\mu}$ in matter~\cite{msw1,msw2,msw3}, 
keeping terms only up to second order in 
the small quantities $\theta_{13}$ and 
$\alpha \equiv \ms/\ma$ \cite{golden,freund}
\be
 P_{e\mu} &\simeq& 
 \sin^2\theta_{23} \sin^22\theta_{13}
\frac{\sin^2[(1-\hat{A})\Delta]}{(1-\hat{A})^2}\nonumber \\
&\pm& \alpha \sin2\theta_{13} \sin2\theta_{12} \sin2\theta_{23} 
\sin\dcp \sin(\Delta) \frac{\sin(\hat{A}\Delta)}{\hat{A}}
\frac{\sin[(1-\hat{A})\Delta]}{(1-\hat{A})} \nonumber \\
&+& \alpha \sin2\theta_{13} \sin2\theta_{12} \sin2\theta_{23} 
\cos\dcp \cos(\Delta) \frac{\sin(\hat{A}\Delta)}{\hat{A}}
\frac{\sin[(1-\hat{A})\Delta]}{(1-\hat{A})} \nonumber \\
&+& \alpha^2 \cos^2\theta_{23} \sin^22\theta_{12} 
\frac{\sin^2(\hat{A}\Delta)}{{\hat{A}}^2}
,
\label{equ:pemu}
\ee
where 
\be
\Delta\equiv \frac{\ma L}{4E},
~~
\hat{A} \equiv \frac{A}{\ma},
~~
 A=\pm 2\sqrt{2}G_FN_eE.
\label{equ:matt}
\ee
Above,  $A$ is the matter potential, expressed 
in terms of the electron density $N_e$ and the 
(anti)neutrino energy $E$; 
the `$+$' sign 
refers to neutrinos while the `$-$' to antineutrinos. 

From Eq. (\ref{equ:pemu}) it is apparent that at a baseline for which
\be
\sin(\hat{A}\Delta) \simeq 0 \, ,
\label{equ:condmagic}
\ee
the oscillation probability is insensitive to $\dcp$. This
defines the ``magic baseline''. As noted, the CERN-INO distance
closely matches this baseline and it has been shown in earlier
work  \cite{paper1,betaino1,betaino2} that this ensures a high
sensitivity at the same time to the determination of
$\theta_{13}$ and the extraction of the neutrino mass hierarchy.
In this work, we examine the benefits which accrue from combining
two Beta-beam experiments both with the source at CERN, one with
the detector at INO and the other located at an appropriate
distance where the impact of CP violation is prominent. Taking a
cue from an optimization analysis of such Beta-beam set-ups
\cite{bboptim}, this second detector is assumed to be at Gran Sasso.

We will refer to  Eqs. (\ref{equ:pemu}) and (\ref{equ:matt}) from 
time to time to
identify the physics underlying the results 
which we present later.

\section{The Experimental Set-up}

The number of (anti)muon events 
in the $i$th energy bin in the detector is given by 
\be
N_{i} = \frac{T\, n_n\, f_{ID}\,\epsilon}{4\pi L^2}~  
\int_0^{E_{\rm max}} dE
\int_{E_{A_i}^{\rm min}}^{E_{A_i}^{\rm max}}
dE_A \,\phi(E) \,\sigma_\numu(E) \,R(E,E_A)\, P_{e\mu}(E) \, ,
\label{equ:events}
\ee
where $T$ is the total running time, 
$n_n$ is the number of target nucleons in the detector,
$f_{ID}$ is the charge identification efficiency 
(meaningful only for the magnetized iron detector), 
$\epsilon$ is the detector efficiency 
and  
$R(E,E_A)$ is the energy resolution function of the detector,
which we assume is a Gaussian. 
For muon (antimuon) events, $\sigma_\numu$ is the 
neutrino (antineutrino) interaction cross-section. 
The quantities $E$ and $E_A$ are the true and 
reconstructed (anti)neutrino 
energies respectively. The quantity $\phi(E)$ is the 
(anti)neutrino Beta-beam flux produced at the source, 
with Lorentz boost $\gamma$ and $L$ is the baseline. 

\begin{table}[t]
\begin{center}
\begin{tabular}{||c||c||c||c||c||c||c||} \hline \hline
   Ion & $\tau$ (s) &
$E_0$ (MeV)
   & $f$& Decay fraction & Beam \\
\hline
  $^{18} _{10}$Ne &   2.41 & 3.92&820.37&92.1\%& $\nu_{e}$    \\
  $^6 _2$He   &   1.17 & 4.02&934.53&100\% &$\bar\nu_{e}$    \\
\hline
 $^{8} _5$B& 1.11 & 14.43&600872.07&100\%&$\nu_{e}$    \\
 $^8 _3$Li& 1.20 &13.47 &425355.16& 100\% & $\bar\nu_{e}$    \\
\hline \hline
\end{tabular}
\caption{\label{tab:ions}
Beta decay parameters: lifetime $\tau$,
electron total end-point energy
$E_0$, $f$-value
and decay fraction for various ions~\cite{beta}. }
\end{center}
\end{table}

\subsection{The Source Ions}

The technological challenge for producing the Beta-beam 
involves creating, bunching, accelerating and storing  
beta unstable 
radioactive ions \cite{lindroos,betabeampage}. 
A pure and intense $\nue$ and/or $\anue$ 
beam is produced when these highly accelerated 
ions decay along the straight sections of the storage ring. 
The choice of ions is governed by a number of factors which 
include their end-point energy, life-time, production rate 
and their charge to mass ratios. Two sets of source ions have 
been identified as plausible candidates for the production of a
Beta-beam. The most widely discussed ions are \neon and \he 
for $\nue$ and $\anue$ beam respectively. The alternative 
set of ions which have also been considered extensively 
in the recent literature are \br and \li for 
$\nue$ and $\anue$ beam respectively. We present 
in Table \ref{tab:ions} the relevant details of the 
properties of these ions. We note that the main difference 
between the two sets lies in their total end-point energies --
\br ($^8$Li) has an end-point energy, $E_0$, which is 3.68 (3.35) times 
that of \neon ($^6$He). The Beta-beam flux spectrum depends solely 
on $\gamma$ and $E_0$ and the energy
of the beam is often crucial in determining the type of physics
that can be explored. We therefore
reiterate the following points \cite{bboptim,issphysics}: %
\begin{itemize}
\item Since the peak neutrino energy is roughly given 
as $\gamma E_0$, the boost factor needed for \neon ($^6$He) 
should be about 3.68 (3.35) times that needed by \br ($^8$Li)
in order to achieve the same neutrino peak energy. 

\item Since the total number of neutrinos peaked in the 
forward direction, {\em i.e.,} towards the detector, roughly goes 
as $\gamma^2$, larger $\gamma$ enhances the flux.
Higher boost factors are therefore preferred. 


\item The reference number of useful decays which we use for
antineutrinos ($2.9\times 10^{18}$ per year) is larger than that
for neutrinos ($1.1 \times 10^{18}$ per year). The ratio between
the two just about compensates for the smaller antineutrino
interaction cross section in matter. So, roughly, this results in a
symmetry between the $\nu$ and $\bar\nu$ data.

\end{itemize}

We have seen in \cite{betaino1,betaino2} that the higher 
end-point energy of \br and \li allows these ions to produce 
Beta-beams with peak energy in the multi-GeV regime, where 
one obtains near-resonant matter effects for the  
near-magic baseline. This can be achieved 
with plausible values of $\gamma$, and results in remarkable 
sensitivity to $\theta_{13}$ and $sgn(\ma)$. Indeed 
this set of ions emerged as the preferred choice 
for the magic baseline after a thorough scan of 
plausible $\gamma$ values \cite{bboptim}. As mentioned 
before, the $\dcp$ sensitivity is smothered at the 
magic baseline. It was seen in \cite{bboptim} that 
intermediate baselines are far superior for CP studies. 
It was further noted  that CP sensitivity 
with \neon and \he is much better than that with \br and $^8$Li. 
Therefore, we take \br and \li as source ions for the 
CERN-INO baseline and \neon and \he as the source ions 
for the CERN-LNGS baseline. We will fix the 
Lorentz boost of \br and \li for CERN-INO as $\gamma=650$ and 
will study the impact of $\gamma$ for \neon and \he at the 
shorter baseline. 

Throughout this work we deal with results obtained by combining
data from neutrino {\em and} antineutrino beams. Further, we
ascribe the same Lorentz boost $\gamma$ to both beams. It has to
be borne in mind  that the
charge to mass ratios, $(Z/A)$, of the source ions are not the
same (see Table \ref{tab:ions}), and so the same boost cannot be
achieved if the ions run simultaneously in a ring. Note, however,
\begin{equation}
\left(\frac{Z}{A}\right)_{^8\rm B}:\left(\frac{Z}{A}\right)_{^{18}\rm Ne} =
\left(\frac{Z}{A}\right)_{^8\rm Li}:\left(\frac{Z}{A}\right)_{^6\rm He} =
9:8
\label{equ:q/m}
\end{equation}
Consequently, if the two neutrino (antineutrino) source ions are
run simultaneously in the ring in which the \br (\li) ions have a
Lorentz boost $\gamma$ = 650, then the \neon (\he) source ions
will have a $\gamma$ of 575. Of course, these are but sample
boost factors for illustrating the feature that simultaneous run
configurations are possible for the chosen source ions, with the
proviso that the (\br and \li) pair have a different value of
$\gamma$ related to that of the (\neon and \he) pair.

\subsection{The ICAL@INO detector}

%
\begin{table}[t]
\begin{center}

\begin{tabular}{||c||c||c||} \hline\hline

\multicolumn{1}{||c||}{{\rule[0mm]{0mm}{6mm}\multirow{2}{*}{Detector Characteristics}}}
& \multicolumn{1}{|c||}{\rule[-3mm]{0mm}{6mm}{ICAL@INO}}
& \multicolumn{1}{|c||}{\rule[-3mm]{0mm}{6mm}{TASD@LNGS}}
\cr
& (Only $\mu^{\pm}$) & (Both $\mu^{\pm}$ \& $e^{\pm}$) \cr
\hline\hline
Total Mass & 50 kton & 50 kton \cr
\hline
Energy threshold & 1 GeV & 0.5 GeV \cr
\hline
Detection Efficiency ($\epsilon$) & 80\% & 80\% ($\mu^{\pm}$) \& 20\% ($e^{\pm}$) \cr
\hline
Charge Identification Efficiency ($f_{ID}$)& 95\% & No CID \cr
\hline
\multirow{2}{*}{Energy Resolution ($\sigma$) (GeV)}& \multirow{2}{*}{0.15E(GeV)} & 0.03$\sqrt{\rm E
(GeV)}$ for $\mu^{\pm}$ \cr
                      & &  0.06$\sqrt{\rm E (GeV)}$ for $e^{\pm}$ \cr
\hline
Bin Size & 1 GeV & 0.2 GeV \cr
\hline
Background Rejection & 0.0001 & 0.001 \cr
\hline
Signal error & 2.5\% & 2.5\% \cr
\hline
Background error & 5\% & 5\% \cr
\hline\hline
\end{tabular}
\caption{\label{tab:detector}
Detector characteristics used in the
simulations. The bin size is kept fixed, while the number of bins
is varied according to the maximum energy.
}
\end{center}
\end{table}

\begin{figure}[t]
\includegraphics[height=7.1cm,width=8.5cm]{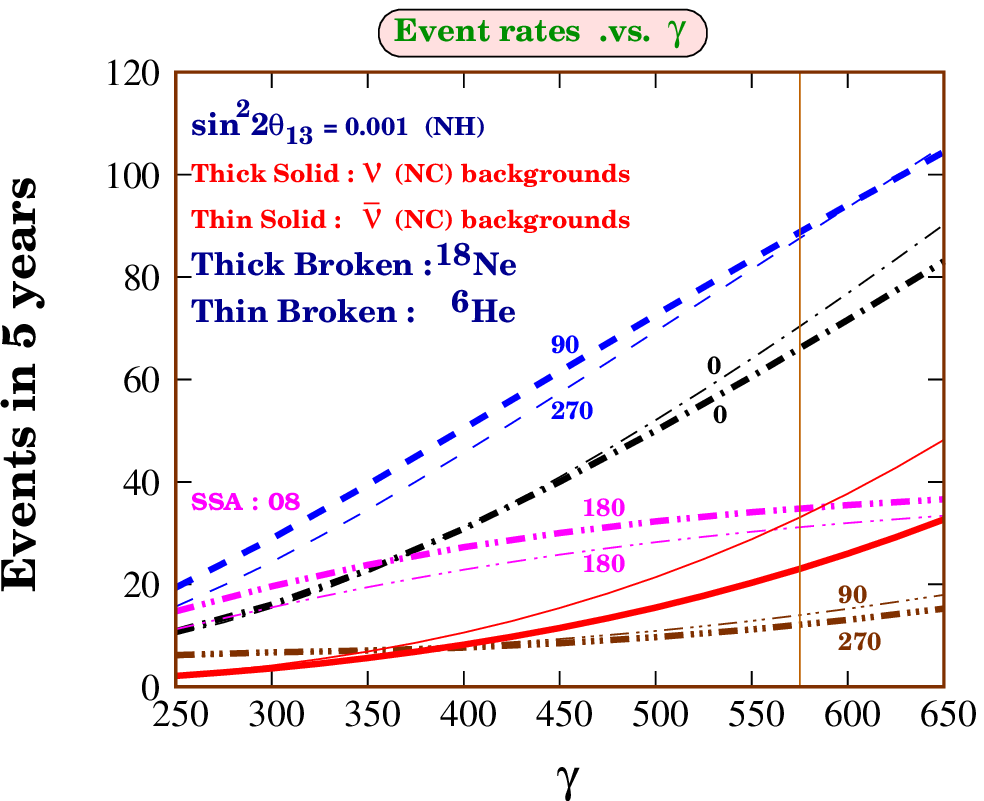}
\vglue -7.1cm \hglue 8.5cm
\includegraphics[width=0.49\textwidth]{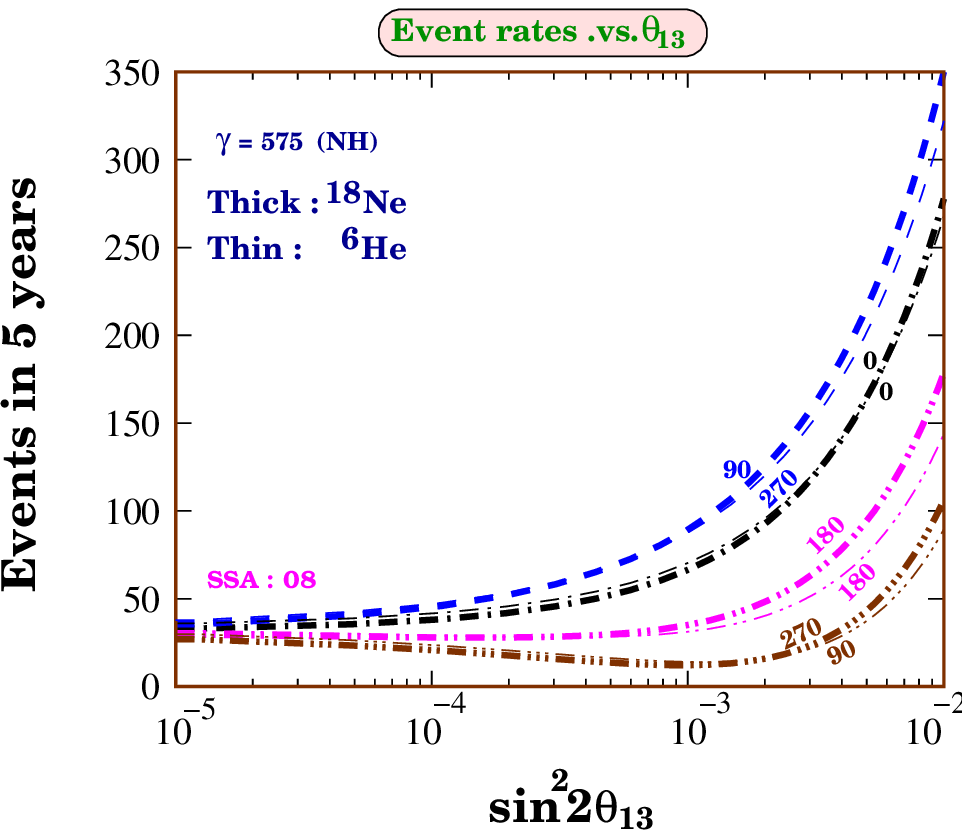}
\caption{\label{fig:event}
Total event rates expected in 5 years at the CERN-TASD@LNGS set-up
assuming $1.1\times 10^{18}$ and $2.9\times 10^{18}$ useful 
decays per year for \neon and \he, respectively. The left panel shows the 
dependence on the boost factor at 
$\stch=10^{-3}$, while the right panel depicts the 
variation with $\stch$ for a fixed $\gamma=500$.
Results are shown for four values of $\dcp$: $0^\circ, 90^\circ,
180^\circ$, and $270^\circ$ for both neutrinos (thick broken lines) and
antineutrinos (thin broken lines). Normal hierarchy has been
assumed. In the left panel, the
vertical line corresponds
to $\gamma = 575$ (see text) and the solid curves are
the effective neutral current backgrounds for neutrinos (thick
red lines)
and antineutrinos (thin red lines).
}
\end{figure}

The ICAL detector at INO will be a 50 kton magnetized iron 
calorimeter with Resistive Plate Chambers (RPCs) serving as the
active elements \cite{ino}. The first phase of this detector
(starting around 2012) will focus on atmospheric neutrino
measurements. 
The entire ICAL detector will be split into 3 modules, each of
which would be of dimension 16 m $\times$ 16 m $\times$ 12 m.
The detector structure consists of iron layers of 6 cm thickness
with a 2 cm gap between layers wherein the glass RPCs will be
interleaved.  The detector would be magnetized by a field of
about 1 Tesla.  The detector characteristics that we have used for
our simulation are shown in Table \ref{tab:detector}. It is planned
that ICAL@INO will be upgraded to 100 kton in the future. 
Number of expected muon (or antimuon) events can be calculated 
using Eq. (\ref{equ:events}).

\subsection{The TASD@LNGS detector}

For the CERN-LNGS baseline we will be working with a 
50 kton TASD. 
A scintillator detector has several
virtues. It has good detection efficiency for muons, can support
a low energy threshold and has excellent energy resolution. It
can also detect electrons but with a lower efficiency. On the
flip side, unlike in the iron calorimeter, there is no charge
identification. 
The detector characteristics
we have assumed \cite{nova,betaoptim}
are listed in Table  \ref{tab:detector}. One can
see from the Table that the background suppression capability of
a TASD is also expected to be 
one order poorer than that of an
iron calorimeter detector\footnote{This can be traced to the
fact that in scintillator detectors hadrons travel farther
before they stop or decay, enhancing the possibility
that they  mimic the signal. The lack of charge identification is
also a handicap in this regard.}. The effect of this on the
results is discussed later. However, the biggest advantage of 
TASD over iron calorimeters is its much lower threshold 
energy. This helps in eliminating 
the clone solutions and hence delivers better CP sensitivity. 

We begin with some remarks about the performance of the
CERN-TASD@LNGS Beta-beam set-up in isolation before turning to
our main theme, namely, a combined analysis of the results from
two detectors at different baselines. The number of useful ion
decays that we choose as reference values are $(1.1 \times
10^{18})$ per year for \neon and $(2.9\times 10^{18})$ per year
for \he. For comparison we also present results
obtained using luminosities one order higher than these reference
values. We combine data from $\nu$ and $\bar\nu$ runs, and assume
that both are at the same Lorentz boost $\gamma$. The values of
the oscillation parameters and the earth's matter density profile
that we use are listed in Table \ref{tab:bench}.

In Fig. \ref{fig:event} the five-year 
(anti)muon event rates for the TASD
detector at Gran Sasso are shown. In the left panel the number of
events is shown as a function of  $\gamma$ for $\stch = 10^{-3}$
while in the right panel it is shown as a function of $\stch$
when $\gamma$ is fixed at 575. In both panels results are shown
for both $\nu$ and $\bar\nu$ beams. Normal neutrino mass
hierarchy has been assumed. 
Notice that in both panels, to a good approximation, the number of
events for neutrinos for any  $\dcp$ closely matches that
for antineutrinos but with CP phase $-\dcp$. This
behavior can be readily understood from Eq. (\ref{equ:pemu}). At
the CERN-LNGS shorter baseline, matter effects are small and hence 
$\hat{A} << 1$ in Eq. (\ref{equ:matt}). It follows that the
$\dcp$ dependence of the probabilities  can be expressed
as $P_{e\mu} \simeq A_0 + A_- \sin \dcp + A_+ \cos
\dcp$ and $P_{\bar{e}\bar{\mu}} \simeq A_0 - A_- \sin
\dcp + A_+ \cos \dcp$, where $A_0, A_\pm$ are
independent of $\dcp$, whence the symmetry is manifest.
In the left panel, the increase in the number of events with
increasing Lorentz boost $\gamma$ is a consequence of the
resultant enhancement of the on-axis flux as well as the higher
energy of the beam. The reason why neutrino 
event rates are higher for 
$\dcp=90^\circ$ and $0^\circ$ compared to that for $\dcp=180^\circ$ 
and $270^\circ$ can be easily seen from the above expressions. 
One can explicitly check that $A_+$ and $A_-$ are positive, 
which means that for $\dcp=180^\circ$ and $270^\circ$ 
number of events are given by a difference of two positive terms 
while for $\dcp=90^\circ$ and $0^\circ$ they come as the corresponding sum.
The reason why the slope of the curves for 
$\dcp=180^\circ$ and 
$\dcp=270^\circ$ ($90^\circ$) for neutrinos (antineutrinos)
is relatively smaller than that of 
$\dcp=0^\circ$ and 
$\dcp=90^\circ$ ($270^\circ$) for neutrinos (antineutrinos)
can be understood from a similar reasoning. 
The behavior of the number of events with $\sin^2 2\theta_{13}$
(right panel) is not difficult to explain. As $\stch$ increases
from a near-zero value, first the 2nd and 3rd terms in  Eq.
(\ref{equ:pemu}), which are linear in $\sin 2\theta_{13}$ and are
dependent on $\dcp$, begin to contribute. Beyond a certain
point, the first term takes over leading to the increasing
behavior for all curves in the right panel.


For the purpose of illustration, in the left panel of Fig.
\ref{fig:event} we also present the $\nu$ and $\bar\nu$ NC
background estimates. Notice that the background is quite 
significant for the entire range of $\gamma$. 
We have also indicated the point
corresponding to $\gamma = 575$, which, as discussed earlier, is
the $\gamma$ for \neon ($^6$He) 
ions when they are run in the same
storage ring as \br ($^8$Li) with 
the $\gamma$ for the latter being 650.

Note that in our analysis 
for this detector, we have also included the
information from the $P_{ee}$ and $P_{\bar e \bar e}$ channels,
albeit with the reduced detector efficiency (see Table
\ref{tab:detector}) of only $20\%$ for
electrons. The number of electron events can be 
calculated using Eq. (\ref{equ:events}), by making appropriate 
changes to the oscillation probability and cross-sections. 
However, as was noted in \cite{dissappear,pee}, 
this channel has hardly any sensitivity to $\theta_{13}$ 
and mass ordering at low baselines like the one under discussion. 
It is independent of $\dcp$.

\begin{table}[t]
\begin{center}
\begin{tabular}{||c||c||} \hline\hline
\multicolumn{1}{||c||}{{\rule[0mm]{0mm}{6mm}{Benchmark Values}}}
& \multicolumn{1}{|c||}{\rule[-3mm]{0mm}{6mm}{$1\sigma$ estimated error}}
\cr
\hline \hline
$|\Delta m^2_{31}{\rm (true)}| = 2.5 \times 10^{-3} \ {\rm eV}^2$ 
& $\sigma(|\Delta m^2_{31}|)=1.5\%$ 
\cr \hline
$\sin^2 2 \theta_{23}{\rm (true)} = 1.0$ 
& $\sigma(\sta)=1\%$ 
\cr \hline
$\Delta m^2_{21}{\rm (true)} = 8.0 \times 10^{-5} \ {\rm eV}^2$ 
& $\sigma(\Delta m^2_{21})=2\%$ 
\cr \hline
$\sin^2\theta_{12}{\rm (true)} = 0.31$ 
& $\sigma(\sss)=6\%$ 
\cr \hline
$\rho{\rm (true)} = 1~{\rm (PREM)}$ & $\sigma(\rho)=5\%$
\cr
\hline \hline
\end{tabular}
\caption{\label{tab:bench}
Chosen benchmark values of the oscillation parameters and the
earth matter density profile and their 
$1\sigma$ estimated error. 
}
\end{center}
\end{table}

\section{CP sensitivity:  CERN-TASD@LNGS Set-up}

The goal of this work is to explore the power of a
two-detector Beta-beam experiment to unravel neutrino mass
parameters. The detector at the magic baseline is insensitive to
the CP phase $\dcp$ and so the main emphasis of the closer
detector will be to address this issue. To set the stage for the
full two-detector analysis,  in this section we discuss the
CERN-TASD@LNGS combination as a stand-alone set-up for exploring
CP violation.

\begin{figure}[t]
\begin{center}
\includegraphics[width=0.49\textwidth]{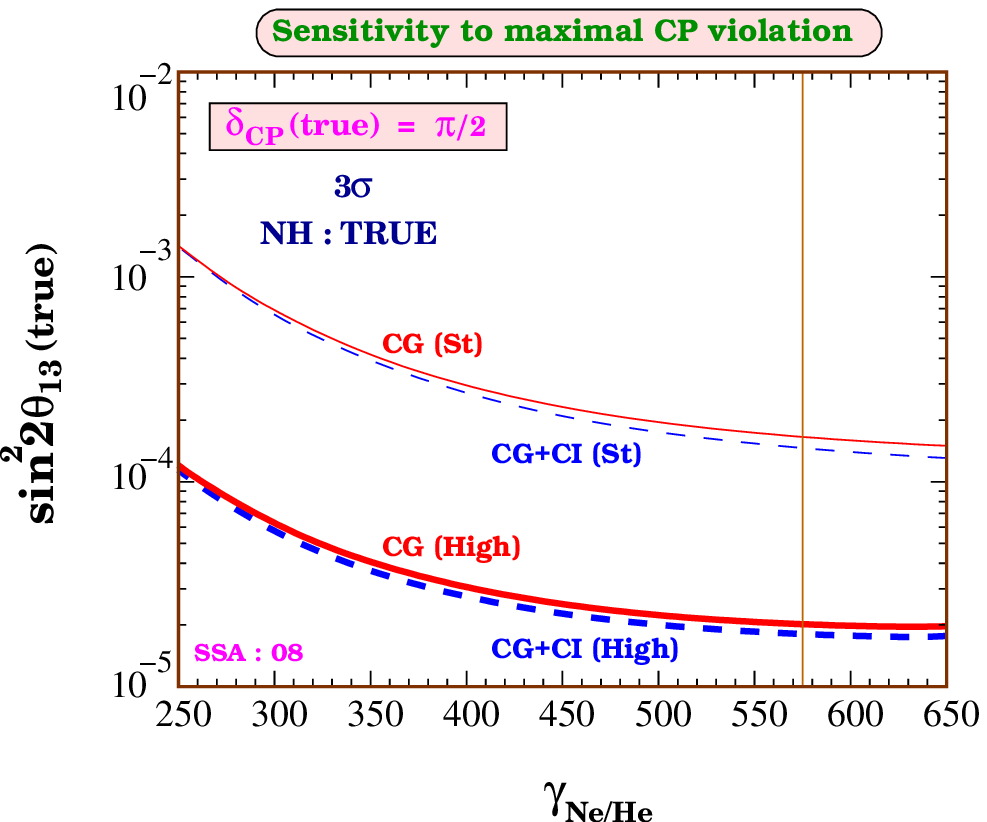}
\includegraphics[width=0.49\textwidth]{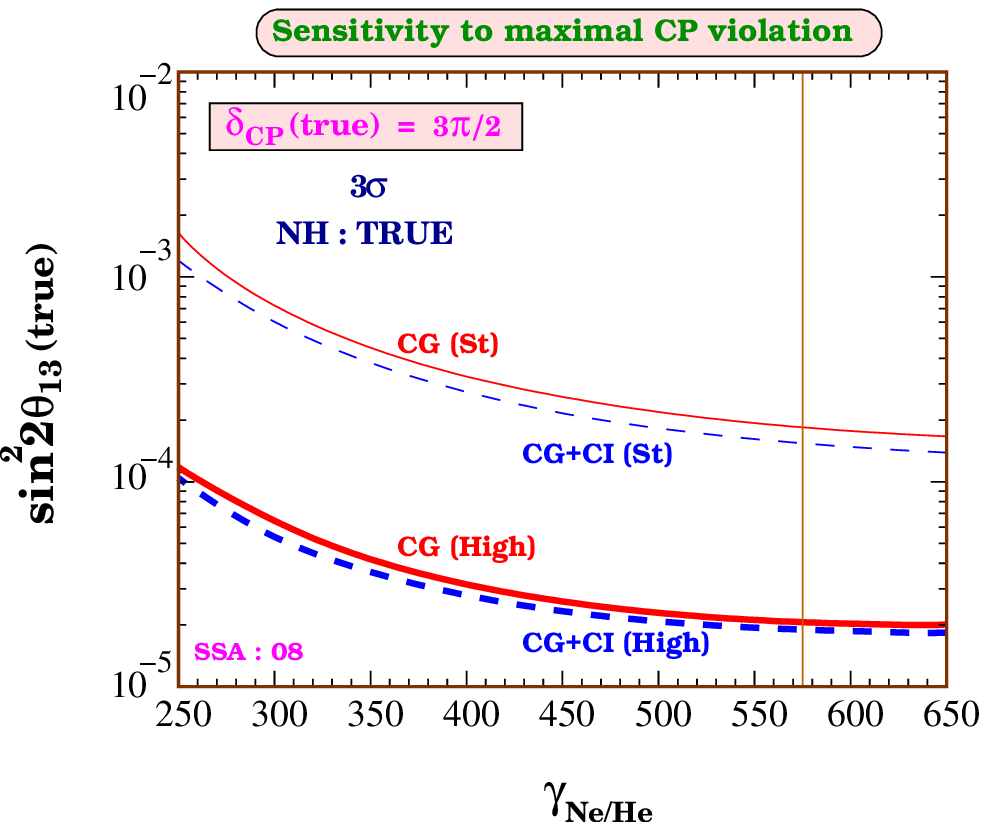}
\caption{\label{fig:maxcp}
The $3\sigma$ $\stcht$ reach for sensitivity to 
``maximal CP violation''. The red solid curves (marked as `CG') 
are for the CERN-TASD@LNGS alone while the blue 
dashed lines (marked as `CG+CI') are for the combined 
data from CERN-TASD@LNGS and CERN-ICAL@INO. 
The results are shown as a function 
of the Lorentz boost for \neon and \he
(taken same for both ions), for 
$\dcpt=90^\circ$ (left panel) and 
$\dcpt=270^\circ$ (right panel). 
Thick lines (marked ``High") are for $5\times  (1.1 \times
10^{19})$ useful \neon and \br decays and $5\times (2.9\times
10^{19})$ useful \he and
\li decays, while thin lines (marked ``St") are for $5\times  (1.1 \times
10^{18})$ and $5\times (2.9\times 10^{18})$ useful ion decays
respectively. 
}
\end{center}
\end{figure}

The $3\sigma$ sensitivity to ``maximal CP violation'' is
presented in Fig. \ref{fig:maxcp} as a function of the boost
factor $\gamma$: these are the (red) solid curves marked ``CG"
in both panels. This performance indicator is defined as follows.
We generate the data for $\dcpt=90^\circ$ (left panel) or
$\dcpt=270^\circ$ (right panel), while in theory we allow
$\dcp$ to be $0^\circ$ and $180^\circ$. We marginalize
over the normalization of the PREM density profile and all other
oscillation parameters, including the neutrino mass hierarchy.
The lowest value of $\stcht$ which allows us to rule out at
$3\sigma$ a CP conserving theory (both $0^\circ$ and $180^\circ$)
when CP is maximally violated in Nature, is plotted in the figure
as the sensitivity reach for maximal CP violation.  


The close similarity of the sensitivity reaches displayed in the
two panels of Fig. \ref{fig:maxcp} is in line with
expectation\footnote{This feature was also noted in
\cite{bboptim} where the detector was a 50 kton magnetized iron
calorimeter.}. From the approximate symmetry
\begin{equation}
P_{e\mu} \longleftrightarrow P_{\bar e \bar\mu} ~~{\rm under }~~
\dcp \longleftrightarrow -\dcp
\label{equ:cpsym}
\end{equation}
noted earlier from Eq. (\ref{equ:pemu}), and recalling that the
data involves $\nu$ as well as $\bar\nu$ beams, it is to be
expected that results for $\dcpt=90^\circ$ and $\dcpt=270^\circ$
will be almost identical\footnote{This symmetrical behavior
breaks down for larger values of $\stcht$. See later.}.  More
explicitly, the statistics for  $\dcpt=270^\circ$ is
comparatively poor for neutrinos with NH true, which is made up
by the higher statistics for antineutrinos.  The effect of the
relatively lower statistics for neutrinos shows up mainly as
additional degeneracies.  As discussed in the previous section,
the better energy resolution and lower threshold of the TASD
results in an amelioration of the correlation and degeneracies
and the islands appearing in Fig. 5 of \cite{bboptim} are absent
here.

We have assumed NH to be the true mass hierarchy in these plots.
This is not expected to be a serious issue as at this baseline
matter effects are small. In fact, the symmetry in  Eq.
(\ref{equ:cpsym}) becomes exact if in addition to $\dcp
\leftrightarrow -\dcp$ the neutrino mass hierarchy is
also flipped. In view of this, for IH true, the left (right) panel of  Fig.
\ref{fig:maxcp} would match with  $\dcpt=270^\circ$ ($\dcpt=90^\circ$).
Of course, this argument rests on Eq. (\ref{equ:pemu}),
which is an approximation, while the results are based on the
exact expressions. We have checked
that indeed taking IH to be true gives very similar results.

It is seen from Fig. \ref{fig:maxcp} that in both panels the
sensitivity reach for maximal CP violation  improves sharply with
$\gamma$ initially, and then it more or less saturates beyond
$\gamma \gtap 500$. This can be understood as follows. For 
$L=730$ km, $P_{e\mu}$ peaks around 1.5 GeV, and 
is very small beyond $\sim 3$ GeV. 
For \neon and \he ions, the peak (anti)neutrino 
energy reaches 1.5 GeV for $\gamma \simeq 430$. For 
$\gamma > 430$, while the peak energy shifts linearly 
to higher values and the total flux increases as $\gamma^2$, 
the flux at the energy $E\simeq 1.5$ where $P_{e\mu}$ is 
high remains 
roughly the same (cf. Fig. 1 of \cite{bboptim}). 
Therefore, for all $\gamma > 430$ we 
do not see any marked improvement. 

In Fig. \ref{fig:maxcp} results are shown for the standard
luminosity as well as for the case where the total statistics is
increased by a factor of ten.  Increasing the luminosity has a
significant impact on the sensitivity of the experiment. From the
figure one can read off that for $\gamma = 575$ and $\dcpt=90^\circ$,
CP violation can be discovered at $3\sigma$ for $\stcht > 1.6 \times
10^{-4}$ for the standard luminosity and for $\stcht > 1.97
\times 10^{-5}$ if luminosity is increased by a factor of ten.
For $\dcpt=270^\circ$, these limits are only
slightly worse, namely, $\stcht > 1.8 \times 10^{-4}$  and
$\stcht > 2.03 \times 10^{-5}$, respectively. Note that the
results are for the NH true case but, as discussed earlier, for
IH true the numbers will hardly change.

\begin{figure}[t]
\begin{center}
\includegraphics[width=0.49\textwidth]{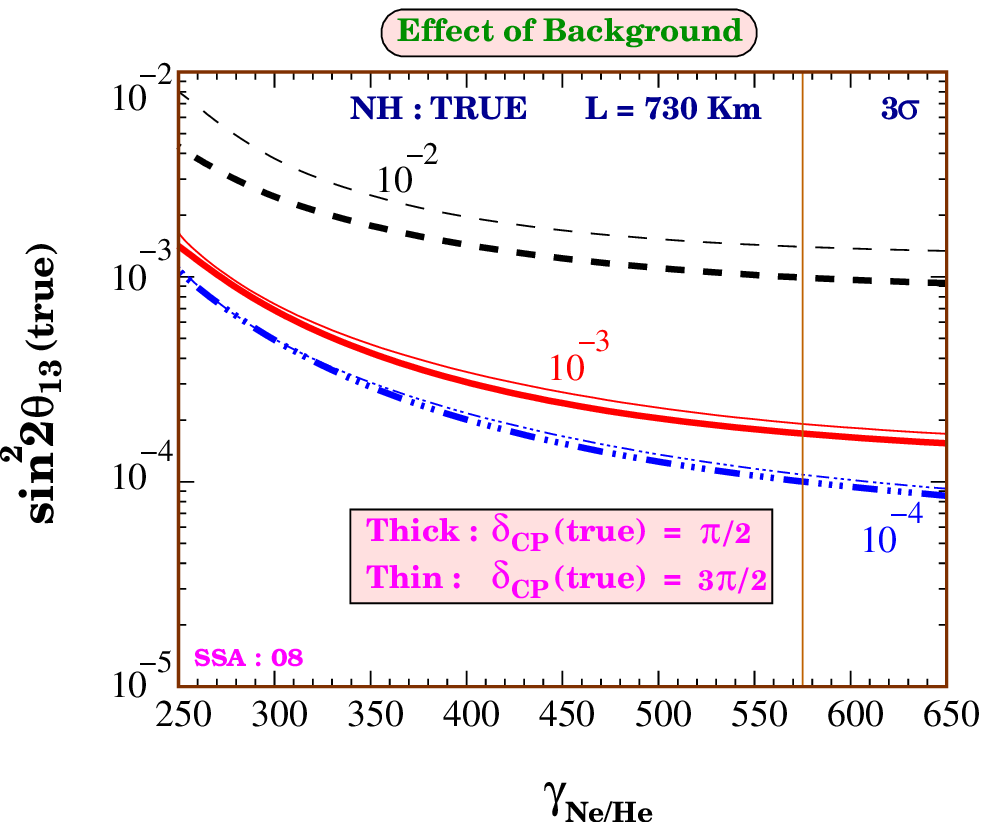}
\includegraphics[width=0.49\textwidth]{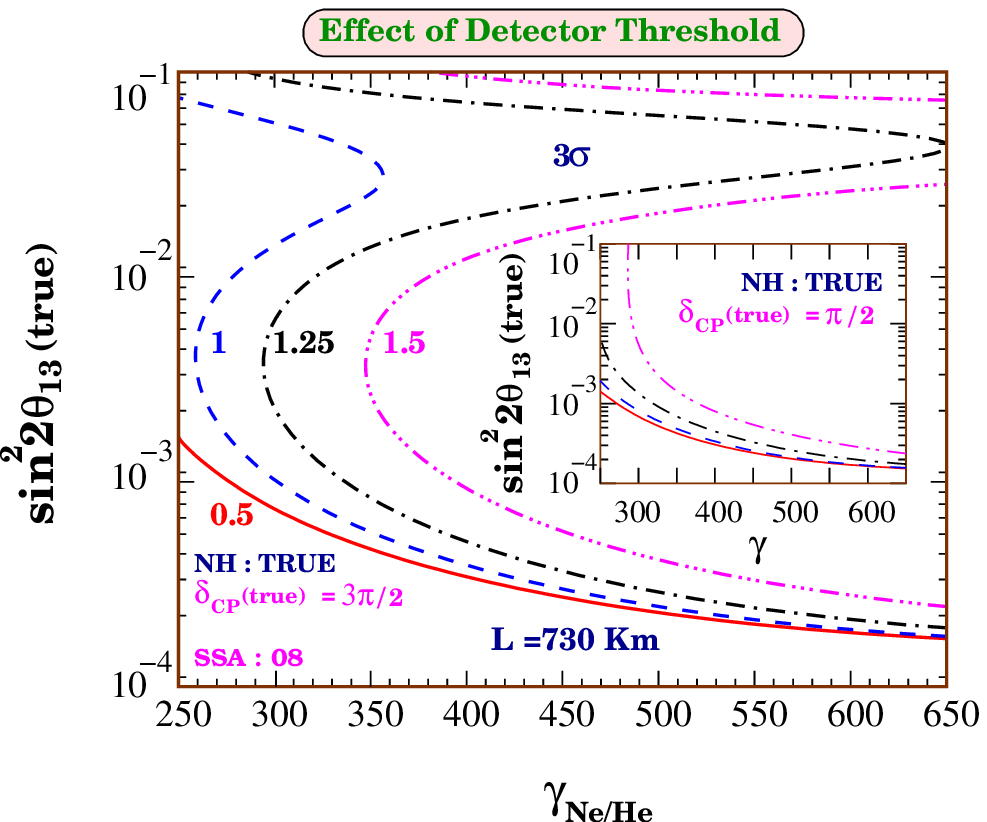}
\caption{\label{fig:tasdcheck}
Effect of changing the detector characteristics on the 
``maximal CP violation'' sensitivity. The left panel
shows the impact of changing the background fraction to one-tenth
or ten times the reference value of $10^{-3}$ (red solid curves).
Results are shown 
for $\dcpt=90^\circ$ by thick lines and 
$\dcpt=270^\circ$ by thin lines.  
The location of $\gamma=575$ is also shown. 
The right panel shows the 
effect on changing the detector threshold. 
Results for 
$\dcpt=90^\circ$ are shown in the inset. The red solid curve
corresponds to a threshold of 500 MeV while the blue dashed,
black-dot-dashed, and pink triple-dot-dashed curves correspond to
thresholds of 1 GeV, 1.25 GeV, and 1.5 GeV, respectively. 
}
\end{center}
\end{figure}

\begin{figure}[t]
\begin{center}
\includegraphics[width=10.0cm]{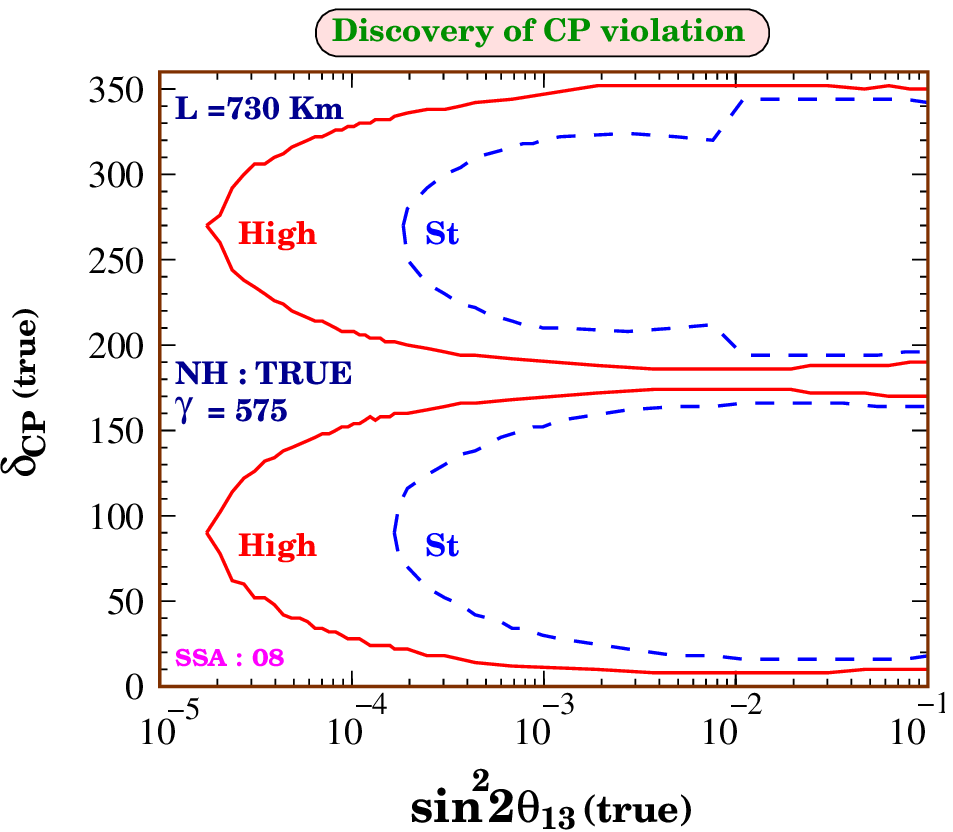}
\caption{\label{fig:cprng}
The area enclosed by the curves represents the $3\sigma$ range of
$\dcpt$ as a function of $\stcht$ for which the data can be used
to rule out the CP-conserving scenario using the CERN-Gran Sasso
reference TASD set-up with
\neon and \he as source ions. The 
Lorentz boost (the same for both ions) is fixed at 575. Results
are shown for  5 years run with the reference luminosity (blue
dashed curves) as well as for a luminosity which is one order
higher (red solid curves).  }
\end{center}
\end{figure}

In Fig. \ref{fig:tasdcheck} we show the impact of changing the
detector characteristics on the maximal CP violation sensitivity
of the TASD@LNGS set-up. The sensitivity is shown in this figure
as a function of $\gamma$ for a 5-year run with the reference
choices of the ($\nu, \bar\nu$) luminosities. 
The left panel shows the effect of 
changing the background fraction from the standard assumed value
of $10^{-3}$ (red solid lines).  We have shown for comparison
results for background fractions of $10^{-2}$ (black dashed
lines) and $10^{-4}$
(blue triple-dotted-dashed lines).  We conclude  (at
$\gamma=575$) that while lowering the background fraction by a
factor of 10 improves the sensitivity by a factor of 1.7, a 10-fold
increase in the background would deteriorate the sensitivity by a
factor of 6. 
Right panel of  the figure shows the
impact of changing the detector threshold on the CP sensitivity.
Increasing the threshold could help us in reducing the
backgrounds\footnote{The atmospheric neutrino background would
be larger at lower energies. Even backgrounds from neutral
current events could be reduced by going to higher threshold.}.
We see that for $\dcpt=270^\circ$ increasing the detector
threshold results in the appearance of degenerate solutions. For
$\dcpt=90^\circ$ (inset) the effect is less severe; however, the
sensitivity falls for both choices of $\dcpt$ as the threshold is
increased. The appearance of clone solutions (multiple solutions
for the same $\gamma$ at different values of  $\stcht$) when the
threshold is increased beyond 1 GeV reflects the important role
of lower energy ({\em i.e.,} lower oscillation wavelength)
neutrinos in this analysis. A strength  of the scintillator
detector is the scope of inclusion of the low energy data 
by virtue of the permissible lower threshold. We 
have also studied the effect of changing the bin size of 
the data on the CP sensitivity of our experimental set-up. 
We find the bin size as large as 1 GeV could also be accommodated
without significantly deteriorating the CP violation sensitivity 
of the experiment.

Fig. \ref{fig:maxcp} gives the ``maximal CP violation"
sensitivity reach, {\em i.e.,} the ability to distinguish $\dcpt$
= 90$^\circ$ or 270$^\circ$ from no CP violation using the
CERN-TASD@LNGS detector. These choices
of $\dcpt$ are the extremal ones. What about other values of $\dcpt$?
In Fig. \ref{fig:cprng} we show for $\gamma$ = 575, the range
of $\dcpt$ as a function of $\stcht$  for which one can
use the data to rule out no CP violation at the 3$\sigma$ level.
We show results with $5\times (1.1
\times 10^{18})$ and $5\times (2.9\times 10^{18})$ useful ion
decays in neutrino and antineutrino modes respectively (the blue
dashed curves) as well as those with one order of magnitude
enhanced number of events in both channels (red solid curves).
At the sensitivity reach values of $\stcht$ displayed in Fig.
\ref{fig:maxcp}, this range is reduced to two points. 
The range grows rapidly as $\stcht$ increases to about 1$\times
10^{-3}$ beyond which it saturates, except for the kinks in the
third and fourth quadrants of  $\dcpt$ around $\stcht$ = 1$\times
10^{-2}$, which arise from clone solutions. For 
$\stcht \geq 10^{-3}$ we should be able to discover CP violation 
for 64\% of the possible $\dcpt$  values 
for the standard luminosity. If luminosity was 
enhanced by a factor of 10, this would improve to 88\% 
of the possible $\dcpt$  values. In fact, for 
10 times larger statistics, CP violation can be 
discovered for 64\% of the possible $\dcpt$  values
for $\theta_{13}$ as low as $\stcht \geq 8\times 10^{-5}$.

\section{Two-baseline combined results}

We are now in a position to discuss the benefits accruing from
combining the results of the two detectors: one at a near-magic
baseline distance and the other located 730 km from the source.
We discuss in turn the sensitivity reaches of this combined
set-up to maximal CP violation, $\sin^2 2\theta_{13}$, and the mass hierarchy.

Recall that the very long baseline detector  is
a magnetized iron calorimeter at the India-based Neutrino
Observatory at a distance of 7152 km from CERN. For this detector
we consider \br and \li as the source ions for $\nu$  and
$\bar\nu$ beams and we assume that both are accelerated to a
reference Lorentz boost $\gamma$ = 650. The reference choice for the number
of useful \br (\li) ion decays  is $5\times  (1.1 \times
10^{18})$ ($5\times (2.9\times 10^{18})$). We also show 
results for a 10 times enhanced luminosity. 

\begin{figure}[t]
\begin{center}
\includegraphics[width=0.49\textwidth]{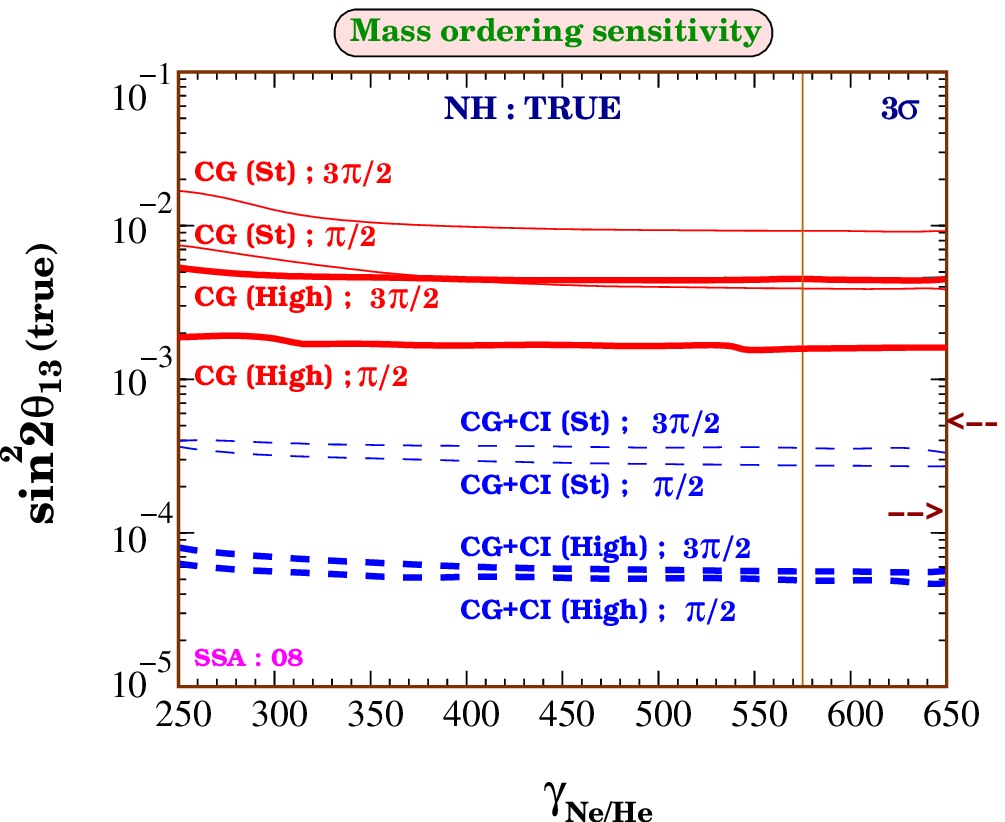}
\includegraphics[width=0.49\textwidth]{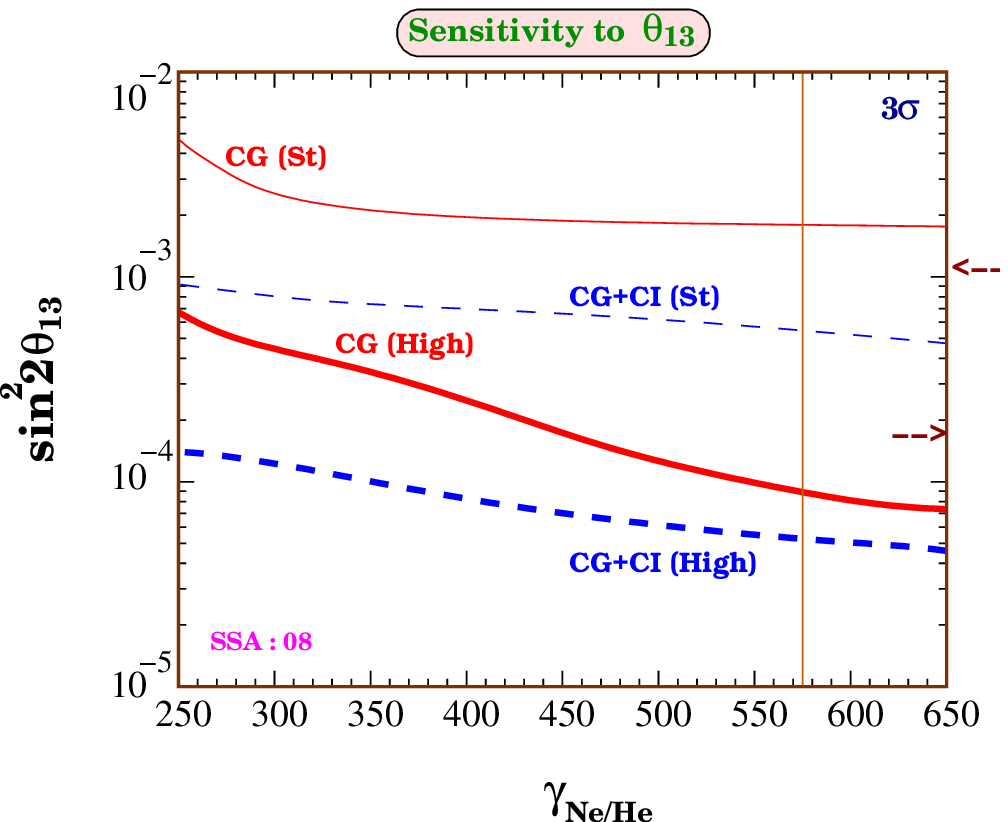}
\caption{\label{fig:sensitivity}
The left and right panels depict  the $sgn(\ma)$ sensitivity
reach and the $\stch$ sensitivity reach, respectively, at
$3\sigma$ as a function of the boost factor for \neon and $^6$He.
In both panels, the red solid lines are for CERN-TASD@LNGS alone
while the blue dashed lines are for the combined data from
CERN-ICAL@INO and CERN-TASD@LNGS. Results for $\dcpt = \pi/2$ and
$3\pi/2$ are shown.  Thick lines are for $5\times (1.1 \times
10^{19})$ useful \neon and \br decays and $5\times (2.9\times
10^{19})$ useful \he and \li decays, while thin lines are for
$5\times  (1.1 \times 10^{18})$ and $5\times (2.9\times 10^{18})$
useful ion decays respectively. The sensitivity reaches for the
CERN-ICAL@INO set-up alone are indicated for both luminosities
(for $\dcpt$ = 0 in the left panel) by arrows on the right side of the panels.  
The location of $\gamma$ = 575 is shown.}
\end{center}
\end{figure}

\subsection{CP Sensitivity of the Combined Set-up}

We first probe the impact of adding the data from the CERN-ICAL@INO
set-up on the sensitivity reach in $\stcht$ for maximal CP
violation. We chose to use the CERN-LNGS baseline with
\neon and \he as source ions since it was shown in \cite{bboptim}
to be the ideal set-up for probing $\dcp$.  In Fig.
\ref{fig:maxcp} is shown (blue dashed curves) the impact of
adding the CERN-ICAL@INO \br and \li results on the sensitivity
to maximal CP violation.  It is seen that there is insignificant
improvement in the sensitivity.  We can conclude that the
CERN-LNGS baseline, which is the optimal choice as a stand-alone
set-up for CP measurements, remains so after the addition of the
new magic baseline data. This should not be 
surprising since at the
magic baseline there is no $\dcp$ dependence, and 
the CP violation measurement of the  
730 km experiment is almost devoid of parameter degeneracies. 

\subsection{The {$\mathbf {sgn(\ma)}$} Sensitivity Reach}

Next, we consider the potential of the two-baseline Beta-beam
set-up to explore the nature of the neutrino mass hierarchy.
The ``$sgn(\ma)$ sensitivity reach'' is defined as the range of values of
$\stcht$ for which the experiment could rule out
the wrong hierarchy at $3\sigma$. We show this as a function of
the boost of the \neon and \he ions in the left panel of Fig.
\ref{fig:sensitivity}.  The boost for the
\br and \li source ions for
CERN-ICAL@INO is fixed at $\gamma=650$. We have shown results for
$\dcpt=90^\circ$ and $270^\circ$ 
and assuming  NH to be true.  The solid lines show
the $sgn(\ma)$ sensitivity of CERN-TASD@LNGS alone while the
dashed lines give the sensitivity when the two data sets are
combined.  
The $sgn(\ma)$ sensitivity reaches for the CERN-ICAL@INO set-up
with $\gamma=650$ with the standard and
enhanced luminosities (for
$\dcpt$ = 0)
are indicated in the figure by  arrows. 
In Table \ref{tab:compare} we give for a fixed $\gamma$, 
the sensitivity reach in $\stcht$ for determining the mass 
hierarchy for the single and combined baseline set-ups 
considered in this paper. A comparison with the sensitivity 
reach envisaged for the optimized two-baseline neutrino factory 
from Ref. \cite{issphysics} is also given for comparison. 
From the figure and the table we note that 
the hierarchy sensitivity is rather poor for the
CERN-TASD@LNGS set-up alone  since the baseline is not large
enough to capture enough matter effect and is almost independent
of the value of $\gamma$. As shown in
\cite{betaino2}, the CERN-ICAL@INO experiment has a very good hierarchy
sensitivity. 
Notice the 
near-independence on $\dcpt$ -- a feature of the magic baseline. 
For the combined two-baseline set-up we 
achieve an exceptional sensitivity such that (for $\dcpt=90^\circ$)
the mass hierarchy could be determined at $3\sigma$ 
if $\stcht > 2.7 \times 10^{-4}$  and $4.64 \times 10^{-5}$
for the standard  and ten times enhanced luminosity respectively.
All results presented are for NH true, 
however, we have checked that results with IH true are similar. 

\subsection{The $\mathbf {\sin^22\theta_{13}}$ Sensitivity Reach}

Finally, we turn to the expected sensitivity of the 
combined set-up to $\theta_{13}$. 
We define the performance indicator 
``$\stch$ sensitivity reach'' as follows. We generate the 
data at $\stcht=0$ and determine the values of 
$\stch$ that could fit this data within a chosen confidence level, 
allowing all oscillation parameters, including $\dcp$ and 
the mass hierarchy, to vary freely in the fit. We also 
marginalize over the normalization of the PREM density 
profile. 
This is applicable when the experiment has observed 
no $\theta_{13}$ driven signal and at best can put an upper limit 
on the still allowed $\stch$. 
The $3\sigma$
projected $\stch$ sensitivity as a function of $\gamma$  for \neon
and \he is shown in the right panel of Fig. \ref{fig:sensitivity}.
The boost of \br and \li for the CERN-ICAL@INO data is fixed at
$\gamma=650$.  The red solid lines in the figure show the $\stch$
sensitivity for CERN-TASD@LNGS alone, while the dashed lines are
what we expect by combining the two data sets. 
The sensitivity reach in $\stch$ for the CERN-ICAL@INO 
is given in Table \ref{tab:compare} and 
is indicated in the figure by an arrow.  
As can be seen from Table \ref{tab:compare}, the 
sensitivity reach for CERN-TASD@LNGS alone is comparable to 
that of CERN-ICAL@INO taken alone. 
For ten times enhanced statistics the corresponding $\stch$
sensitivity reach of CERN-ICAL@INO and CERN-TASD@LNGS 
taken alone are, $\stch <
1.76 \times 10^{-4}$ and $\stch < 8.59 \times 10^{-5}$
respectively.  A combination of the two data sets improves the
performance tremendously.  For the standard luminosity we could
reach down to $\stch < 5.46 \times 10^{-4}$, which could be
further improved to $\stch < 5.26 \times 10^{-5}$ in the event
that ten
times larger statistics becomes feasible.


\begin{table}[p]
\begin{center}
{\footnotesize
\begin{tabular}{|c||c|c||c|c||c|c|} \hline\hline
\multirow{4}{*}{Set-up}
& \multicolumn{2}{|c||}{{\rule[0mm]{0mm}{6mm}Mass Ordering (\sig)}}
& \multicolumn{2}{|c||}{\rule[-3mm]{0mm}{6mm}{CP Sensitivity (\sig)}}
& \multicolumn{2}{|c|}{\rule[-3mm]{0mm}{6mm}{$\stch$ Sensitivity (\sig)}} \cr
& \multicolumn{2}{|c||}{{\rule[0mm]{0mm}{3mm}NH (True)}}
& \multicolumn{2}{|c||}{{\rule[0mm]{0mm}{3mm}NH (True)}}
& \multicolumn{2}{|c|}{{\rule[0mm]{0mm}{3mm}}} 
\cr \cline{2-7} 
&  $1.1\times 10^{18}$ & $1.1\times 10^{19}$ &
   $1.1\times 10^{18}$ & $1.1\times 10^{19}$ &
 $1.1\times 10^{18}$ & $1.1\times 10^{19}$ \cr
&  \& & \& &
   \& & \& &
   \& & \& \cr
&  $2.9\times 10^{18}$ & $2.9\times 10^{19}$ &
   $2.9\times 10^{18}$ & $2.9\times 10^{19}$ &
 $2.9\times 10^{18}$ & $2.9\times 10^{19}$ \cr

\hline\hline
CERN-INO & $4.7\times 10^{-4}$ & $9.4\times 10^{-5}$ &
 Not & Not & 
 \multirow{2}{*}{$1.14\times 10^{-3}$} & \multirow{2}{*}{$1.76\times 10^{-4}$} \cr
$\gamma = 650$, 7152 Km  & ($4.9\times 10^{-4}$) & ($1.2\times 10^{-4}$) &
  possible & possible & 
  & \cr

\hline\hline
CERN-LNGS & $3.89\times 10^{-3}$ & $1.58\times 10^{-3}$ &
 $1.6\times 10^{-4}$ & $1.97\times 10^{-5}$ &
 \multirow{2}{*}{$1.78\times 10^{-3}$} & \multirow{2}{*}{$8.59\times 10^{-5}$} \cr
$\gamma = 575$, 730 Km  & ($9.23\times 10^{-3}$) & ($4.48\times 10^{-3}$) &
  $(1.8\times 10^{-4})$ & $(2.03\times 10^{-5})$ &
  & \cr

\hline\hline
CERN-BOULBY & $2.49\times 10^{-3}$ & $2.19\times 10^{-4}$ &
 $1.85\times 10^{-4}$ & $1.99\times 10^{-5}$ &
 \multirow{2}{*}{$1.41\times 10^{-3}$} & \multirow{2}{*}{$1.45\times 10^{-4}$} \cr
$\gamma = 575$, 1050 Km  & ($7.87\times 10^{-3}$) & ($4.1\times 10^{-3}$) &
  $(2.02\times 10^{-4})$ & $(2.04\times 10^{-5})$ &
  & \cr

\hline\hline
CERN-LNGS & \multirow{3}{*}{$2.7\times 10^{-4}$} & \multirow{3}{*}{$4.64\times 10^{-5}$} &
 \multirow{3}{*}{$1.42\times 10^{-4}$} & \multirow{3}{*}{$1.78\times 10^{-5}$} &
 \multirow{4}{*}{$5.46\times 10^{-4}$} & \multirow{4}{*}{$5.26\times 10^{-5}$} \cr
 $\gamma = 575$, 730 Km & & &
  & &
  & \cr
$+$ & & &
  & &
  & \cr
CERN-INO & \multirow{-3}{*}{($3.58\times 10^{-4}$)} & \multirow{-3}{*}{($5.45\times 10^{-5}$)} &
 \multirow{-3}{*}{($1.49\times 10^{-4}$)} & \multirow{-3}{*}{($1.88\times 10^{-5}$)} &
 & \cr
 $\gamma = 650$, 7152 Km & & &
  & &
  & \cr

\hline\hline
CERN-BOULBY & \multirow{3}{*}{$2.67\times 10^{-4}$} & \multirow{3}{*}{$4.57\times 10^{-5}$} &
 \multirow{3}{*}{$1.63\times 10^{-4}$} & \multirow{3}{*}{$1.8\times 10^{-5}$} &
 \multirow{4}{*}{$6.1\times 10^{-4}$} & \multirow{4}{*}{$6.69\times 10^{-5}$} \cr
 $\gamma = 575$, 1050 Km & & &
  & &
  & \cr
$+$ & & &
  & &
  & \cr
CERN-INO & \multirow{-3}{*}{($3.37\times 10^{-4}$)} & \multirow{-3}{*}{($5.17\times 10^{-5}$)} &
 \multirow{-3}{*}{($1.76\times 10^{-4}$)} & \multirow{-3}{*}{($1.87\times 10^{-5}$)} &
 & \cr
 $\gamma = 650$, 7152 Km & & &
  & &
  & \cr

\hline\hline
\multicolumn{7}{|c|}{\rule[0mm]{0mm}{6mm}{Optimized Neutrino Factory
set-up with two improved golden detectors (50 kton each) placed at}} \cr
\multicolumn{7}{|c|}{\rule[0mm]{0mm}{6mm}
{$L$ = 4000 km \& 7500 km respectively. $E_{\mu} = 20$~GeV \& total $5 \times 
10^{21}$ decays for $\mu^{-}$ \& $\mu^{+}$ each.}}  
\cr 
\hline\hline
Optimized
& \multicolumn{2}{|c||}{{\rule[0mm]{0mm}{6mm}$4.5\times 10^{-5}$}}
& \multicolumn{2}{|c||}{\rule[-3mm]{0mm}{6mm}\multirow{2}{*}{$1.5\times 
10^{-5}$}}
&
\multicolumn{2}{|c|}{\rule[-3mm]{0mm}{6mm}\multirow{2}{*}{$4.5\times
10^{-5}$}} \cr Neutrino Factory &
\multicolumn{2}{|c||}{{\rule[0mm]{0mm}{3mm}(100\% of $\dcpt$
coverage)}}
& \multicolumn{2}{|c||}{{\rule[0mm]{0mm}{3mm}}}
& \multicolumn{2}{|c|}{{\rule[0mm]{0mm}{3mm}}}
\cr
\hline\hline
\end{tabular}
}
\caption{\label{tab:compare}
Performances of various Beta-beam set-ups at $3\sigma$
in addressing the key unsolved issues: mass ordering, CP violation and
$\stch$ sensitivity reach. For CP sensitivity and mass ordering,
the minimum values of $\stcht$ required for a positive conclusion
are presented. Results are shown for a five-year run with the
reference
luminosity: $1.1 \times 10^{18}$ ($2.9\times 10^{18}$)
useful ion decays per year
in the $\nu$ ($\bar\nu$) mode as well as one order of magnitude
higher statistics. The numbers without (with)  parantheses
correspond to $\dcpt=90^\circ$ ($\dcpt=270^\circ$).
Note that the $\stch$ sensitivity reach is independent of the value of $\dcpt$
and the true mass ordering because the
prospective ``data'' have been generated
at $\theta_{13}$ = 0. The CERN-INO baseline is insensitive to
$\dcp$. For comparison, the expectations from an optimized
two-baseline Neutrino Factory set-up with upgraded
magnetized iron detectors are also listed \cite{issphysics,nufactoptim}.
}

\end{center}
\end{table}

\section{Discussions and Conclusions}

In previous papers \cite{betaino1,betaino2} we have shown that 
a Beta-beam experiment where $\nue$ and $\anue$ beams 
produced using \br and \li ions respectively 
are sent from CERN to the
magnetized iron calorimeter ICAL at INO
results in tremendous sensitivity to $\theta_{13}$ and 
the mass hierarchy. This unprecedented sensitivity stems from 
a combined effect of the CERN-INO distance being 
magic, as well as the energy of the Beta-beam being conducive 
to encountering near-maximal matter effects, thereby 
resulting in substantial enhancement of the oscillation 
probability and hence statistics. However, 
while the baseline 
being magic and the beam energy being in the multi-GeV regime 
is a virtue for probing $\theta_{13}$ and 
the mass hierarchy, it becomes a serious drawback for 
CP violation studies. This necessitates the deployment of a 
second baseline option with a lower energy neutrino beam for 
investigating all the three issues simultaneously. The optimal 
baseline and ion source option for CP violation studies was 
shown in \cite{bboptim} to be $L=600-900$ km and 
the \neon and \he combination with intermediate values of $\gamma$. 
In this paper we studied the sensitivity reach of 
the combined data sets from the CERN to INO magical Beta-beam 
set-up and a CERN to Gran Sasso Beta-beam experiment with 
\neon and \he as source ions. We reiterate that even for 
the Neutrino Factory experiment, 
at least two baselines are needed in order 
to optimally address all the three neutrino oscillation 
parameters. 

For the CERN to INO set-up, 
the baseline is $L=7152$ km and we used \br and \li as sources, 
with boost $\gamma=650$. The far detector was taken as 
ICAL, a 50 kton magnetized iron calorimeter with 
detection efficiency of 80\%, charge identification efficency 
of 95\%, energy threshold of 1 GeV, 
energy resolution of $0.15 E$ and neutral current background 
fraction of $10^{-4}$. We call this set-up CERN-ICAL@INO. 
For the CERN to Gran Sasso sector, 
the baseline is $L=730$ km and we chose \neon and \he as source ions. 
The Lorentz boosts for the \neon and \he ions are taken
to be the same and  allowed to vary between 
250 and 650. Because of the lower beam energy for this case  
it is preferable to opt for a Totally Active Scintillator 
Detector here. We assume an active detector 
mass of 50 kton with 80\% detection efficiency for 
muons and 20\% detection efficiency for electrons, 
energy threshold of 0.5 GeV, 
energy resolution of $0.03 \sqrt{E({\rm GeV})}$ for muons and 
 $0.06 \sqrt{E({\rm GeV})}$ for electrons  
and neutral current background 
fraction of $10^{-3}$. We call this set-up CERN-TASD@LNGS. 
We show results for ``standard luminosity'' where we use 
the above-mentioned detector sizes and efficiencies and 
$5\times (1.1 \times 10^{18})$  
and $5\times (2.9\times 10^{18})$ 
useful ion decays 
in neutrino and antineutrino modes respectively. 
We also show results for a situation where the statistics is  
ten times larger. 

We probed the physics potential of this two-detector set-up 
with a Beta-beam as the neutrino source. We have presented our 
results in terms of three performance indicators.
The essential results are summarized in Table \ref{tab:compare}. 
For $\dcp$ sensitivity we showed the $\stcht$ reach 
of the experiment to distinguish maximal CP violation from 
a CP-conserving scenario. We displayed the  results for  
the CERN-TASD@LNGS experiment by itself and when data from here are combined 
with that from CERN-ICAL@INO. From CERN-TASD@LNGS alone we 
obtain magnificent sensitivity to $\dcp$. For NH true and 
$\gamma=575$ for \neon and \he and  with the standard luminosity,  
maximal CP violation can be established at $3\sigma$ at CERN-TASD@LNGS 
if $\stcht > 1.6 \times 10^{-4}$ for
$\dcpt=90^\circ$. 
With ten times more statistics this 
limit gets pushed to  $\stcht > 1.97 \times 10^{-5}$. Addition of 
the CERN-ICAL@INO data does not improve these limits significantly.
We also explored the dependence of maximal CP violation sensitivity 
of CERN-TASD@LNGS to the detector characteristics of TASD. Results 
are similar if IH is true. 

Measurement of $\theta_{13}$ and mass hierarchy can be performed 
extremely well at CERN-ICAL@INO. We studied the impact of adding the 
data from CERN-TASD@LNGS on the final combined sensitivity to these 
parameters. 
For the mixing angle $\theta_{13}$ we defined the $\stch$ 
sensitivity as the range of $\stch$ which 
could fit at $3\sigma$ the data generated for $\stcht=0$, after 
marginalization over all oscillation parameters and the density 
profile. We found that the 
$\stch$ sensitivity of both set-ups taken alone are comparable. 
CERN-ICAL@INO could limit $\stch < 1.14 \times 10^{-3}$ 
for the standard luminosity and with $\gamma=650$, while 
CERN-TASD@LNGS could restrict $\stch < 1.78 \times 10^{-3}$ 
with $\gamma=575$. For the enhanced luminosity these 
limits would be $\stch < 1.76 \times 10^{-4}$ and 
$\stch < 8.59 \times 10^{-5}$ respectively. 
However, when added together, the combined data can limit 
$\stch < 5.46 \times 10^{-4}$ for the standard luminosity and 
$\stch < 5.26 \times 10^{-5}$ with a factor of ten enhanced luminosity.  

The sensitivity to $sgn(\ma)$ is defined in terms of the 
$\stcht$ required in order to rule out the wrong 
hierarchy at $3\sigma$. We had shown earlier that for 
CERN-ICAL@INO with $\gamma=650$ and standard luminosity 
the wrong inverted hierarchy can be disfavored at $3\sigma$ if 
$\stcht > 4.7 \times 10^{-4}$ for $\dcpt=90^\circ$. 
With ten times more statistics 
this improves to $\stcht > 9.4 \times 10^{-5}$. 
We found in this paper that the $sgn(\ma)$ sensitivity of 
CERN-TASD@LNGS alone is rather poor in comparison. However, 
when we add the two data sets, we find an enhancement in the 
sensitivity. With data from both set-ups taken together 
the wrong inverted hierarchy can be disfavored at  $3\sigma$ if 
$\stcht > 2.7 \times 10^{-4}$ with the standard luminosity 
when $\dcpt=90^\circ$. 
With ten times more statistics 
it would become possible if $\stcht > 4.64 \times 10^{-5}$. 
Results for $\dcpt = 270^\circ$ are similar, and so are the 
sensitivity reaches for IH true. 

Another choice for an intermediate baseline 
could be sending a beam from CERN to the Boulby mine in 
UK. The CERN to Boulby distance is about 1050 km. 
Since INO and Boulby mine are in opposite 
hemispheres with respect to CERN, it might be easier 
to contemplate a decay ring design which could be used to 
send beams to both these (plausible) detector sites.   
While in the text we explicitly presented and discussed 
results for the CERN-LNGS baseline, Table \ref{tab:compare} 
shows results for the CERN-Boulby mine set-up as well. 
All detector characteristics are taken to be the same 
as for the LNGS case. The sensitivity reach can be 
quantitatively seen to be comparable for both the 
intermediate baseline 
options when combined with CERN-INO data, 
with the CERN-LNGS baseline being marginally better. 

The sensitivity obtained in the set-ups considered here could be 
compared to that obtained in a benchmark high $\gamma$ set-up 
considered in \cite{bc2}. In this paper, the authors proposed 
using a $\gamma=350$ $^{18}$Ne and $^6$He Beta-beam sent 
to a megaton water detector located at a distance of 
730 km from CERN. For total exposure of 5.0 Mton-year, 
this set-up returns a $3\sigma$ 
$\stheta$ sensitivity 
of $5.7 \times 10^{-4}$. Normal mass hierarchy  
discovery reach at $3\sigma$ ranges between 
$2.4 \times 10^{-3}$ and $1.6 \times 10^{-2}$ depending on 
the choice of $\dcpt$. Maximal CP violation can be 
established at $3\sigma$ 
if $\stcht \geq 5.2 \times 10^{-5}$ for $\dcpt=\pi/2$ 
and $\stcht \geq 5.5 \times 10^{-5}$ for $\dcpt=3\pi/2$.

The Beta-beam set-ups discussed here have a reach comparable to
those obtained from experiments based on optimised Neutrino
Factories, see Table \ref{tab:compare}. Here, the reference
Neutrino Factory set-up we have considered is the version with
two ``golden magnetized detectors'' of 50 kton each at $L$ = 4000
km and 7500 km, a detector threshold of 1 GeV, 5 years
$\mu^+/\mu^-$ data with $E_\mu$ = 20 GeV and a luminosity of
$10^{21}$ decays per year for each polarity, with a background
fraction of $5 \times 10^{-6}$. Further details of these set-ups
and their resolutions and efficiencies can be found in
\cite{issphysics,nufactoptim}.

It is of interest to compare the sensitivity reach of our
two-detector Beta-beam set-up with that chosen in
\cite{newdonini}, where the authors used $L=2000$ km for the
shorter baseline magnetized iron calorimeter detector
in addition to another 
at the magic baseline, and \br and \li ions as sources for both.
While they use a smaller
boost factor of 350 for the beam, they require larger number of
ion decays than we have used for our standard luminosity.  Our
results with standard luminosity are comparable to those of
\cite{newdonini} with larger luminosity, as noted
in the Introduction.  Similar reach for CP sensitivity in the two
proposals even though with different source ions and intermediate
baseline is not surprising.  The relative optimization between
$\gamma$ and luminosity for the two sets of sources was studied
in detail in \cite{bboptim}. It was shown that to obtain the same
``physics'' with $^8$B/$^8$Li and $^{18}$Ne/$^6$He, one needs 12
times larger luminosity for the former and 3.5 times larger
$\gamma$ for the latter.  While we have used a factor of 1.86
higher boost than that in \cite{newdonini}, the authors of
\cite{newdonini} consider 7.39 (2.80) times higher luminosity for
their neutrino (antineutrino) mode.  For lower number of ion
decays the sensitivity of the set-up proposed in
\cite{newdonini} becomes worse than our set-up.  The main
strength of the set-up proposed here emerges from the fact that 
use has been made of the optimal choice for the source ions and baselines
\cite{bboptim} for CP, $sgn(\ma)$ and $\theta_{13}$.  The boost
factors that we use, though larger, should be plausible at CERN
and were optimized for the CERN-INO set-up for determining
$sgn(\ma)$ and $\theta_{13}$ \cite{betaino2}.  For the
corresponding boost possible for \neon and \he we achieve good
sensitivity to CP violation in the CERN-LNGS set-up.  The
combined data set from both detectors provide exceptional
sensitivity to all three neutrino parameters.  With ten times
larger luminosities, which are also considered to be plausible
\cite{mats_talk_RAL_2008}, the sensitivity of our set-up
escalates to unprecedented levels and are similar to those
possible with a high performance two-detector Neutrino Factory
set-up, using improved detectors with very low backgrounds and
low threshold (see Table \ref{tab:compare}).

In conclusion, we show that a Beta-beam experiment with two
carefully chosen detectors at optimal distances can provide
unprecedented sensitivity to the establishment of CP-violation in
the lepton sector, addressing the issue of neutrino mass
ordering, and determination of the mixing angle $\theta_{13}$. We
have considered a Totally Active Scintillator Detector at the
shorter baseline of 730 km and a magnetized iron calorimeter at a
near-magic baseline distance of 7152 km, both of mass 50 kton. In
tandem, they provide an exceptional sensitivity to the 
neutrino parameters and match the precision achievable in
high performance Neutrino Factory set-ups. 


\vglue 0.8cm
\noindent
{\Large{\bf Acknowledgments}}\vglue 0.3cm
\noindent
The authors acknowledge the HRI cluster facilities for computation.
This work has been supported 
under the XI Plan Neutrino Project at the Harish-Chandra Research Institute.

\setcounter{equation}{0}
\setcounter{section}{0}

\renewcommand{\thesection}{\Alph{section}}
\renewcommand{\theequation}{\thesection-\arabic{equation}}

\section{Appendix: Background Rejection}

In this Appendix we discuss some technical issues pertaining to
rejection of the beam related  and atmospheric neutrino backgrounds.

{\subsection{Beam related neutral current backgrounds}}


From Fig. 1, one can see that at the CERN-LNGS baseline ($L$ = 730 Km), 
the expected event rate sharply depends on the choice of $\dcp$ 
and $\theta_{13}$. Assuming $1.1\times 10^{18}$ useful decays per year for 
\neon and at $\stch = 10^{-3}$ with a boost factor of 575 and 
normal hierarchy, the variation in the signal in 5 years with $\dcp$
has been depicted in Table \ref{tab:signal}. The beam related neutral
current background in this case (after folding with $10^{-3}$ 
suppression factor for TASD) comes out to be 23 events. So in this 
case, the background over signal ratio may vary over a wide range
from 0.26 to 1.92. 
This case is quite different to that 
encountered in the CERN-INO Beta-beam experiment \cite{betaino1,betaino2}
where we hardly expect any beam related background. 
Though the total
signal to neutral current 
background ratio is high, we stress that an aspect which is extremely 
crucial is the different spectral shape of the signal compared to the 
background. 
In our
analysis we have cautiously taken into account the beam related background
as described in the appendix of \cite{betaino2}.


\begin{table}[h]
\begin{center}
\begin{tabular}{||c||c||c||c||c||} \hline \hline
   $\dcp$ & 0 & $90^\circ$ & $180^\circ$ & $270^\circ$ \cr
\hline
   Signal (5 yrs) & $67$ & $90$ & $35$ & $12$ \cr
\hline \hline
\end{tabular}
\caption{\label{tab:signal}
Number of $\nu$ signal events in 5 years for various $\dcp$ 
 with a 50 kton TASD detector
($\stch = 10^{-3}$ and NH with $\gamma$ = 575).  }
\end{center}
\vskip -0.6cm
\end{table}


{\subsection{Atmospheric backgrounds}}

Next we turn to the backgrounds due to
atmospheric neutrinos.
 In the CERN-LNGS set-up, we have considered a TASD
detector with an energy threshold of 500 MeV. With \neon and $\gamma$ = 575,
we have an energy spectrum extending upto 3.9 GeV with a peak at 2.25 GeV.
In this energy range the number of muons from atmospheric neutrinos 
giving tracks inside the detector along the direction of the $\beta$-beam
flux are expected to be very large (see Table \ref{tab:atm}). Without demur
one can see that the atmospheric backgrounds are huge compared to the
signal rates. Also the TASD detector does not have any charge
identification (CID) capability and it makes the case more
arduous in the sense that we have to consider both $\mu^{-}$ and
$\mu^{+}$ atmospheric events simultaneously as the backgrounds
when we expect a signal from the neutrino beam alone.  In such a
situation, the timing information of the ion bunches inside the
storage ring turns out to be a valuable tool to tackle these
backgrounds.


\begin{table}[ht]
\begin{center}
\begin{tabular}{||c||c||c||c||c||} \hline \hline
   Energy range (GeV) & $\mu^{-}$ events & $\mu^{+}$ events \cr
\hline
   0.5 - 4 & 1342 & 447 \cr
\hline
   1 - 4 & 852 & 344 \cr 
\hline \hline
\end{tabular}
\caption{\label{tab:atm}
Expected atmospheric neutrino events along the beam 
direction in 5 years with a 
50 kton TASD detector. 
}
\end{center}
\vskip -0.6cm
\end{table}


{\subsection{Bunch size and background reduction}}


For a 5T magnetic field and $\gamma$ = 575 for \neon ions, the radius of the
curved section would be $R \sim 720~m$. The useful decays are those
which occur in one of the straight sections where emitted neutrinos fly 
towards the direction of the detector. If $L_r$ is the length of one of the 
straight sections of the decay ring and if we demand that 
$L_r/(2 \pi R + 2L_r) \sim 36\%$ (useful fraction of ion decays) 
then $L_r$ comes out to be $5818~m$. 
With this design, the total length of the storage ring ($2 \pi R + 2L_r$) 
would be $16161~m$. Now if we can tolerate the effective atmospheric 
background over beam signal ratio to upto $10^{-1}$ then
\begin{equation}
10^{-1} = \frac {Atmospheric~Background \times Suppression~Factor 
~(S_{f})}{Beam~Signal}
\label{equ:supp}
\end{equation}
The total atmospheric background in the energy range of 0.5 to 4 GeV 
is 1789 (see Table \ref{tab:atm})
and the expected lowest signal rate in 5 years 
from the \neon $\beta$-beam is  
12 (with $\stch = 10^{-3}$ and $\dcp = 270^\circ$).
So from Eq. \ref{equ:supp}, we have $S_{f} = 6.7 \times 10^{-4}$. 
To achieve this $S_{f}$, one has to ensure an excellent
timing information 
for the signal correlating with the source bunches. This
requirement puts a constraint on the maximum allowed time-length of the 
ion bunch $T_{b}$ inside the storage ring. The $S_{f}$ and $T_{b}$ are 
related in the following fashion,
\begin{equation}
S_{f} = \frac {v \times T_{b} \times N_{b}}{2 \pi R + 2L_r}\,\times 0.36
\label{equ:bunch}
\end{equation}
where $v \simeq c$ is the velocity of the ion and $N_{b}$ is the maximum
number of bunches of the ion circulating inside the storage ring at the
same time. One can readily see from Eq. \ref{equ:bunch} that 
$S_{f} = 6.7 \times 10^{-4}$ can be achieved with 12.5 ns time-length 
of the ion bunch taking $N_{b}= 8$.

If we increase the threshold energy of the TASD detector from 0.5
to 1 GeV then we can suppress the atmospheric events by a factor of 1.5 
(see Table \ref{tab:atm}) and now the total atmospheric background would be
1196. But in this process, the signal rate also gets reduced from 12 to 8,
keeping the $S_{f}$ almost same. 

Another important point one should keep in mind is that if we can work with 
one order higher luminosity of the Beta-beam flux then the
signal rate will be enhanced by a factor of 10 and therefore one can
relax the time length of the ion bunch from 12.5 ns to 125 ns provided
that $N_{b}$ is kept fixed.


\end{document}